\documentclass[a4paper,11pt]{article}

\usepackage{jcappub}
\usepackage{slashed}
\usepackage{bm}
\usepackage{latexsym,amssymb,amsmath,float,url,mathrsfs}
\usepackage{tikz}
\usetikzlibrary{tikzmark}
\usepackage{latexsym}
\usepackage{epstopdf}
\usepackage{amsfonts}
\usepackage{amsmath}
\usepackage{amssymb}
\usepackage{comment}
\usepackage{subfigure}
\usepackage{natbib}
\usepackage{hyperref}
\usepackage{pifont}
\usepackage{blindtext}
\usepackage{adjustbox}
\usepackage{multirow}
\usepackage{tabularx}
\usepackage{tikz}
\usetikzlibrary{tikzmark}
\usepackage{amssymb}
\usepackage{bbold}
\usepackage{blkarray}
\usepackage{graphicx}
\usepackage{amsfonts}
\usepackage{soul}
\usepackage{amssymb}
\usepackage{amsmath}
\usepackage{cancel}
\usepackage{tcolorbox}

\usepackage{graphics,appendix,afterpage,makecell} 

\def\Im{\mbox{Im}\,}
\def\Re{\mbox{Re}\,}

\definecolor{oucrimsonred}{rgb}{0.6, 0.0, 0.0}
\definecolor{persianblue}{rgb}{0.11, 0.22, 0.73}
\definecolor{forestgreen}{rgb}{0.13,0.35,0.13}
\definecolor{lightgray}{rgb}{0.83, 0.83, 0.83}
 \hypersetup{colorlinks, citecolor=oucrimsonred, linkcolor=black, urlcolor=oucrimsonred}
\definecolor{cornellred}{rgb}{0.7, 0.11, 0.11}
\definecolor{navyblue}{rgb}{0.0, 0.0, 0.5}
\definecolor{amethyst}{rgb}{0.6, 0.4, 0.8}
\definecolor{yellow}{rgb}{1.0, 1.0, 0.0}
\definecolor{firebrick}{rgb}{0.7, 0.13, 0.13}
\definecolor{tangerineyellow}{rgb}{1.0, 0.8, 0.0}
\definecolor{deepfuchsia}{rgb}{0.76, 0.33, 0.76}
\definecolor{amber}{rgb}{1.0, 0.75, 0.0}
\definecolor{VioletRed4}{rgb}{0.55, 0.13, .32}
\definecolor{indiagreen}{rgb}{0.07, 0.53, 0.03}
\definecolor{VioletRed4}{rgb}{0.55, 0.13, .32}
\newcommand{\be}{\begin{equation}}
\newcommand{\ee}{\end{equation}}
\newcommand{\bea}{\begin{equation} \begin{aligned}}
\newcommand{\eea}{\end{aligned} \end{equation}}

\definecolor{oucrimsonred}{rgb}{0.6, 0.0, 0.0}
\newcommand\vertarrowbox[3][6ex]{%
  \begin{array}[t]{@{}c@{}} #2 \\
  \left\uparrow\vcenter{\hrule height #1}\right.\kern-\nulldelimiterspace\\
  \makebox[0pt]{\scriptsize#3}
  \end{array}%
}
\hypersetup{
     colorlinks   = true,
     citecolor    = violet,
     urlcolor     = violet,
     linkcolor    = violet}

\definecolor{verdechiaro}{rgb}{0.6,1,0.6}
\definecolor{giallochiaro}{rgb}{1,1,0.6}
\definecolor{bluscuro}{rgb}{0.15, 0.2, 0.9}
\definecolor{verdes}{rgb}{0.1, 0.5, 0.1}%
\definecolor{tangerineyellow}{rgb}{1.0, 0.8, 0.0}

\definecolor{americanrose}{rgb}{1.0, 0.01, 0.24}
\definecolor{cobalt}{rgb}{0.0, 0.28, 0.67}
\definecolor{brandeisblue}{rgb}{0.0, 0.44, 1.0}
\definecolor{mycolor}{rgb}{0.0, 0.0, 0.5}
\definecolor{oxfordblue}{rgb}{0.0, 0.13, 0.28}
\definecolor{azure}{rgb}{0.0, 0.5, 1.0}
\definecolor{turquoiseblue}{rgb}{0.0, 1.0, 0.94}
\newtcolorbox{mynewbox}[1]{colback=white!5!white,colframe=azure!75!black,fonttitle=\bfseries,title=#1}
\newtcolorbox{mybox}{colback=mycolor!5!white,colframe=azure!75!black}
\newtcolorbox{mynamedbox}[1]{colback=mycolor!5!white,colframe=azure!75!black,title=#1}
\definecolor{venetianred}{rgb}{0.78, 0.03, 0.08}
\newtcolorbox{mynamedbox1}[1]{colback=venetianred!5!white,colframe=venetianred!80!black,title=#1}
\newtcolorbox{mynamedbox2}[1]{colback=azure!5!white,colframe=azure!80!black,title=#1}

\definecolor{verdes}{rgb}{0.1, 0.5, 0.1}%
\definecolor{cornellred}{rgb}{0.7, 0.11, 0.11}

\definecolor{VioletRed4}{rgb}{0.55, 0.13, .32}

\definecolor{rossocorsa}{rgb}{0.83, 0.0, 0.0}

\def\Im{\mbox{Im}\,}
\def\Re{\mbox{Re}\,}

\renewcommand{\r}{r_s}

\def\l2{\langle\!\langle}
\def\r2{\rangle\!\rangle}

\definecolor{oucrimsonred}{rgb}{0.6, 0.0, 0.0}
\definecolor{persianblue}{rgb}{0.11, 0.22, 0.73}
\definecolor{forestgreen}{rgb}{0.13,0.35,0.13}
\definecolor{lightgray}{rgb}{0.83, 0.83, 0.83}
 \hypersetup{colorlinks, citecolor=oucrimsonred, linkcolor=black, urlcolor=oucrimsonred}
\definecolor{cornellred}{rgb}{0.7, 0.11, 0.11}
\definecolor{navyblue}{rgb}{0.0, 0.0, 0.5}
\definecolor{amethyst}{rgb}{0.6, 0.4, 0.8}
\definecolor{yellow}{rgb}{1.0, 1.0, 0.0}
\definecolor{firebrick}{rgb}{0.7, 0.13, 0.13}
\definecolor{tangerineyellow}{rgb}{1.0, 0.8, 0.0}
\definecolor{deepfuchsia}{rgb}{0.76, 0.33, 0.76}
\definecolor{amber}{rgb}{1.0, 0.75, 0.0}
\definecolor{VioletRed4}{rgb}{0.55, 0.13, .32}
\definecolor{indiagreen}{rgb}{0.07, 0.53, 0.03}
\definecolor{VioletRed4}{rgb}{0.55, 0.13, .32}

\definecolor{oucrimsonred}{rgb}{0.6, 0.0, 0.0}

\definecolor{verdechiaro}{rgb}{0.6,1,0.6}
\definecolor{giallochiaro}{rgb}{1,1,0.6}
\definecolor{bluscuro}{rgb}{0.15, 0.2, 0.9}
\definecolor{verdes}{rgb}{0.1, 0.5, 0.1}%
\definecolor{tangerineyellow}{rgb}{1.0, 0.8, 0.0}

\definecolor{americanrose}{rgb}{1.0, 0.01, 0.24}
\definecolor{cobalt}{rgb}{0.0, 0.28, 0.67}
\definecolor{brandeisblue}{rgb}{0.0, 0.44, 1.0}
\definecolor{mycolor}{rgb}{0.0, 0.0, 0.5}
\definecolor{oxfordblue}{rgb}{0.0, 0.13, 0.28}
\definecolor{azure}{rgb}{0.0, 0.5, 1.0}
\definecolor{turquoiseblue}{rgb}{0.0, 1.0, 0.94}

\definecolor{verdes}{rgb}{0.1, 0.5, 0.1}%
\definecolor{cornellred}{rgb}{0.7, 0.11, 0.11}

\definecolor{VioletRed4}{rgb}{0.55, 0.13, .32}

\hypersetup{
     colorlinks   = true,
     citecolor    = violet,
     urlcolor     = violet,
     linkcolor    = violet}

\def\d{{\rm d}}
\definecolor{rossocorsa}{rgb}{0.83, 0.0, 0.0}

\usepackage[normalem]{ulem}

\usepackage[font=small,labelfont=bf]{caption}
\title{ Schwarzschild Black Hole Turbulence: Scalar Probe }
\author[a]{Alex Kehagias}
\affiliation[a]{Physics Division, National Technical University of Athens, Athens, 15780, Greece}

\author[b]{Antonio Riotto}
\affiliation[b]{Department of Theoretical Physics and Gravitational Wave Science Center,  \\
24 quai E. Ansermet, CH-1211 Geneva 4, Switzerland}
\emailAdd{kehagias@mail.ntua.gr, antonio.riotto@unige.ch }

\abstract{
We explore how  perturbations of a Schwarzschild black hole can redistribute energy among scalar modes and seed turbulent-like cascades. We make use of the  van der Pol–Krylov–Bogoliubov averaging method and derive coupled mode equations that describe near-resonant interactions between neighboring multipoles.  We compare two routes to instability, namely the difference-frequency mixing between adjacent modes and the diagonal (Mathieu) self-modulation channel. We show that, at high multipole number (eikonal limit), the difference-frequency route dominates and drives a one-way cascade from higher to lower frequencies. We chart the corresponding instability regions (“tongues”) and  quantify their detuning dependence. The framework provides a simple, quantitative mechanism for energy transfer in black hole ringdowns and clarifies when and how turbulent signatures can arise within linear probes on a weakly perturbed background.

}

\begin{document}
\maketitle
\flushbottom

\section{Introduction}
Quasi-Normal Modes (QNMs) constitute a defining feature of Black Holes (BHs), capturing their characteristic response to external disturbances (for a recent review, see Ref.~\cite{Berti:2025hly}). By analyzing this response, one can extract valuable insights into the final stage of the merger between compact objects—the so-called ringdown—which marks the birth of a new BH.
Accurately describing the ringdown phase is a fundamental aspect of the ongoing effort to interpret gravitational-wave observations from compact binary coalescences~\cite{LIGOScientific:2020tif,LIGOScientific:2021sio,Capano:2021etf,Finch:2022ynt,Isi:2022mhy,Cotesta:2022pci,Siegel:2023lxl}, as well as to prepare for analyses involving next-generation ground- and space-based detectors~\cite{Berti:2005ys,Ota:2019bzl,Bhagwat:2019dtm,Pitte:2024zbi}.
Although the current framework has yielded considerable success with existing data, several recent works~\cite{London:2014cma,Mitman:2022qdl,Cheung:2022rbm,Ma:2022wpv,Redondo-Yuste:2023seq,Cheung:2023vki,Zhu:2024rej} have emphasized that nonlinear gravitational effects can alter the expected QNM spectrum. In particular, these effects can introduce additional spectral peaks located at frequencies corresponding to linear frequency sums, often with appreciable amplitudes.
A range of analytical and numerical investigations has been devoted to exploring these nonlinear oscillations~\cite{Kehagias:2023ctr,Redondo-Yuste:2023seq,Cheung:2023vki,Perrone:2023jzq,Zhu:2024rej,Ma:2024qcv,Bourg:2024jme,Bucciotti:2024zyp,Khera:2024yrk,Kehagias:2024sgh,Bucciotti:2025rxa,bourg2025quadraticquasinormalmodesnull,BenAchour:2024skv,Kehagias:2025xzm,Ling:2025wfv,Kehagias:2025tqi,Perrone:2025zhy,Fransen:2025cgv,Kehagias:2025ntm,Kehagias:2025gvk,Ianniccari:2025avm,Singh:2025xzd}. These studies mainly focus on the quadratic quasi-normal mode $2\ell$ arising from the coupling of two $\ell$-multipoles, which could potentially be observable with future third-generation detectors for $\ell=2$~\cite{Lagos:2024ekd,Yi_2024,PhysRevD.109.064075,shi2024detectabilityresolvabilityquasinormalmodes}. 

However, how about the opposite process in which two large  $\ell$-modes annihilate to create a smaller  $\ell$-mode? Being of the interaction being of the same (cubic) order, one would expect that this process is equally taking place. This would be an inverse energy cascade, eventually terminating into the most observable mode $\ell=2$.
In fact, recently it has been pointed out by numerical experiments that nonlinear interactions  can also  induce 
inverse energy cascades by inducing resonant  instabilities which  
 transfer energy from higher
 to lower frequencies \cite{Ma:2021znq}. Originally, this phenomenon has been discovered analytically  for rapidly-spinning BHs which can  display turbulent gravitational behavior  mediated by a  parametric instability \cite{Yang:2014tla} and led to several studies regarding turbulence behaviour in gravitational wave physics \cite{Galtier:2017mve,Galtier:2018vbq,Benomio:2024lev,Iuliano:2024ogr,Figueras:2023ihz,Krynicki:2025fzi,Siemonsen:2025fne}.

In this paper we take a step further towards the study of the turbulent behaviour of QNMs, analizying the  case of the   Schwarzschild BH. In particular,  we investigate how small disturbances of a Schwarzschild BH  redistribute energy across scalar modes and can initiate turbulence-like cascades. By applying the so-called van der Pol–Krylov–Bogoliubov averaging technique, we obtain a set of coupled mode equations capturing near-resonant couplings between neighboring angular multipoles. Our analysis contrasts two pathways to instability: mixing at the difference frequency between adjacent modes, and a diagonal self-modulation mechanism akin to a Mathieu-type resonance. In the limit of large multipole indices (the eikonal regime), we will find that the difference-frequency mechanism prevails, producing a unidirectional flow of energy from higher to lower frequencies. 

The paper is organized as follows. In section 2 we describe the dynamics of a massless scalar field subjct to a monochromatic driver and study the two-mode reduction in section 3. Section 4 is dedicated to the higher-order instability. Finally, section 5 offers our conclusions.

\section{Massless scalar on Perturbed Schwarzschild}
Let us consider a massless scalar field propagating in a Schwarzschild spacetime, perturbed by an even–parity, axisymmetric quadrupole fluctuation (Zerilli field)  of the  metric. We choose a scalar field to avoid problems with gauge invariance arising beyond linear perturbation theory. The latter  is a  time-periodic field that acts as a ``pump'' or ``driver'' (to borrow the terminology from nonlinear optics 
\cite{SalehTeich2007,Boyd:2008eba}) and  modulates the  parameters of the system while  transfers energy to secondary (daughter or ``idler'')  modes from the primary (the parent or the ``signal'') mode. As we will see below, the dynamics is of a time–dependent system that exhibits parametric resonance between a parent scalar mode of angular index $\ell$ and a daughter mode of index $\ell-2$, driven by the  $L=2$, $M=0$ gravitational pump at frequency $\Omega$. In particular, we will expand the scalar in QNM radial functions, and by projecting the perturbed wave equation onto the QNM basis, we will isolate the two–mode truncation that exhibits the  resonance tongue. 

The Schwarzschild metric is written in standard Schwarzschild coordinates as 
\begin{eqnarray}
{\rm d} s^2=g^{(0)}_{\mu\nu}{\rm d} x^\mu {\rm d} x^\nu =-f\,{\rm d} t^2+f^{-1}{\rm d} r^2+r^2  \left(\d \theta^2+\sin^2\theta\, \d \phi^2\right),\qquad 
f(r)=1-\frac{2M}{r}. 
\end{eqnarray}
The perturbed metric is then $g_{\mu\nu}=g^{(0)}_{\mu\nu}+ \varepsilon h_{\mu\nu}$, where $\varepsilon\ll 1$ is a bookeeping small parameter and the
the even–parity, axisymmetric quadrupolar perturbation $h_{\mu\nu}$ is taken in Regge–Wheeler gauge to 
be 
\begin{eqnarray}
    h_{\mu\nu}=\left(\begin{array}{cccc}
    f\,H_0&H_1&0&0\\ 
H_1&f^{-1} H_2&0&0  \\
         0&0&r^2K&0\\
         0&0&0&r^2 \sin^2\theta \,K
    \end{array}
    \right) \, Y_{20}(\theta,\phi). 
\end{eqnarray}
The functions $H_0(r,t),H_1(r,t),H_2(r,t),K(r,t)$  are determined algebraically by the Zerilli master field $Z_{20}(t,r)$. We consider a monochromatic QNM driver, with QNM frequency $\Omega$
$$Z_{20}(t,r)=\Re\big[\mathcal Z(r)\,e^{-i\Omega t}\big], $$
and, hence, each of $H_i$ and $K$ are of the same form, i.e.,
\begin{eqnarray}
    H_i(t,r)=\Re\big[\hat H_i(r)\,e^{-i\Omega t}\big], \qquad 
    K(t,r)=\Re\big[\hat K(r)\,e^{-i\Omega t}\big].
\end{eqnarray}
The equation now for a massless scalar $\Phi$ on the perturbed Schwarzschild spacetime with metric  $g_{\mu\nu}=g^{(0)}_{\mu\nu}+ \varepsilon h_{\mu\nu}$ is written as 
\begin{equation}
\Box_{g}\Phi=
\Box_{g^{(0)}}\Phi-\varepsilon h^{ab}\nabla_a\nabla_b\Phi-\varepsilon\left(\nabla^a h_a{}^{\,b}-\frac{1}{2}\nabla^b h\right)\nabla_b\Phi=0,
\label{eq:massles}
\end{equation}
where indices are raised with $g^{(0)}$ and $h=h^a{}_a$. 

Let $\{u_{\ell n}(r)\}$ denote the standard Schwarzschild scalar QNMs with azimuthal angular number $\ell$ and overtone $n$. They  solve the equation
\begin{equation}
\partial_{r_*}^2u_{\ell m}+\Big{(}\omega_{\ell n}^2-V_\ell(r)\Big{)}u_{\ell m}=0
\end{equation}
with the usual ingoing and outgoing boundary conditions at horizon and spatial infinity, respectively,
\begin{eqnarray}
   && u_{\ell n}\sim e^{-i\omega_{\ell n} r_*}, \qquad (r_*\to-\infty),\nonumber \\
   &&
u_{\ell n}\sim e^{+i\omega_{\ell n} r_*}, \qquad (r_*\to+\infty), 
\label{eq:bc}
\end{eqnarray}
whereas the scalar potential is 
\begin{equation}
V_\ell(r)=f\!\left(\frac{\ell(\ell+1)}{r^2}+\frac{2M}{r^3}\right).
\label{eq:potential}
\end{equation}
Although QNMs are non-renormalizable \cite{Kokkotas:1999bd}, following Green-Hollands-Zimmerman (GHZ) \cite{Green:2022htq,Cannizzaro:2023jle}, we may introduce in the  the space of (radial) solutions a conserved, symmetric bilinear form $\l2\cdot,\cdot\r2
$ which is defined by a contour-renormalized Wronskian pairing such that
\begin{equation}
\l2 u_{\ell n},\,u_{\ell' n'}\r2=\mathcal{N}_{\ell n}\,\delta_{\ell\ell'}\,\delta_{nn'},
\qquad
\mathcal{N}_{\ell n}\neq 0,
\label{eq:orth}
\end{equation}
and the QNM excitation coefficients are precisely the projections with respect to $\l2,\r2$. Specifically, one may take
\begin{equation}
\l2u,v\r2
=\int_{\Gamma} u(r_*)\,v(r_*)\,dr_*,
\label{eq:GHZ}
\end{equation}
where the integration path $\Gamma$ is rotated into the complex $r_*$-plane so that the QNM asymptotics are integrable \cite{Green:2022htq,Cannizzaro:2023jle}. In the following, we will only need the orthogonality of the QNMs with respect to the GHZ bilinear form and not the particular representation of the integral in (\ref{eq:GHZ}).  

We proceed by 
expanding the scalar $\Phi$ in the spherical harmonic basis as 
\begin{eqnarray}
\Phi(t,r,\theta,\phi)=\sum_{\ell m}\frac{\psi_{\ell m}(t,r)}{r}\,Y_{\ell m}(\theta,\phi).
\end{eqnarray}
We may now express the mode functions 
$\psi_{\ell m}(t,r)$ in terms of the 
QNMs of the corresponding problem on the exact (unperturbed) Schwarzschild spacetime. 
 Keeping, for simplicity, only the fundamental overtone $n=0$ in each $\ell$, we may write
\begin{equation}
\psi_{\ell m}(t,r)=A_{\ell m}(t)\,u_\ell(r).
\label{eq:QNM-expansion}
\end{equation}
Then, the massless scalar equation (\ref{eq:massles}) on the  perturbed 
Schwarzschild spacetime, after separating  the sperical harmonics  $Y_{\ell m}$, multiplying  with $u_\ell$ and  GHZ-project, is written as  
\begin{eqnarray}
&&\l2u_\ell,\,u_\ell\r2\ddot A_{\ell m}
+\l2u_\ell,\,\omega_\ell^2 u_\ell\r2A_{\ell m}
+\varepsilon\sum_{\ell' m'} \l2u_\ell,\;\mathcal{D}_{\ell\ell'}[h(t),u_{\ell'}]\r2 \mathcal{A}^{(20;0)}_{\ell m;\ell' m'}\,A_{\ell'm'}=0.
\label{eq:amp}
\end{eqnarray}
Here $\mathcal{D}_{\ell\ell'}$ is the radial operator proportional to $\varepsilon$ in Eq. (\ref{eq:massles}). 
In addition, $\mathcal{A}^{(20;0)}_{\ell m;\ell' m'}$ is the purely angular Wigner factor for the even-parity, axisymmetric $L=2,M=0$ driver,
\begin{equation}
\mathcal{A}^{(20;0)}_{\ell m;\ell' m'}
=
\sqrt{\frac{5(2\ell+1)(2\ell'+1)}{4\pi}}\,
\begin{pmatrix}\ell&2&\ell'\\0&0&0\end{pmatrix}
\begin{pmatrix}\ell&2&\ell'\\-m&0&m'\end{pmatrix},
\label{eq:Wigner-GHZ}
\end{equation}
so that axisymmetry enforces $m'=m$ and $\ell'=\ell,\ell\pm2$.
Using \eqref{eq:orth} and defining 
$\mathcal{N}_\ell=\l2 u_\ell,u_\ell\r2$, 
we find,  after dividing \eqref{eq:amp} by $\mathcal{N}_\ell$, the equation
\begin{equation}
\ddot A_{\ell m}+\omega_\ell^2 A_{\ell m}
+\varepsilon\sum_{\ell'm'} \mathcal{M}_{\ell m;\ell'm'}(t)\,A_{\ell'm'}=0,
\label{eq:amp-GHZ-final}
\end{equation}
where the coupling coefficients are 
\begin{equation}
\mathcal{M}_{\ell m;\ell'm'}(t)
=\frac{\l2u_\ell,\;\mathcal{D}_{\ell\ell'}[h(t),u_{\ell'}]\r2}
{\mathcal{N}_\ell}\;
\mathcal{A}^{(20;0)}_{\ell m;\ell' m'}.
\label{eq:GHZ-M-def}
\end{equation}
We may further define the quantities
\begin{equation}
q_\ell=\frac{\l2u_\ell,\;\mathcal{D}_{\ell\ell}[\hat H_i,\,u_\ell]\r2}
{\mathcal{N}_\ell}\;G_{\ell\ell},
\qquad
c_\ell^{\pm}=\frac{\l2 u_\ell,\;\mathcal{D}_{\ell,\ell\pm2}[\hat H_i,u_{\ell\pm2}]\r2}
{\mathcal{N}_\ell}\;G_{\ell,\ell\pm2},
\label{eq:q-c-GHZ}
\end{equation}
where 
\begin{eqnarray}
G_{\ell\ell'}=\mathcal{A}^{(20;0)}_{\ell 0;\ell' 0}=\sqrt{\frac{5(2\ell+1)(2\ell'+1)}{4\pi}}\begin{pmatrix}\ell&2&\ell'\\0&0&0\end{pmatrix}^{\!2}. 
\end{eqnarray}
Then, 
 the projected chain 
 turns out to be
\begin{eqnarray}
&&\ddot A_\ell+\omega_\ell^2 A_\ell
+\varepsilon\,q_\ell\,e^{\Omega_I t}\cos(\Omega_R t)\,A_\ell
+\nonumber \\
&&+\varepsilon\,c_\ell^{+}\,e^{\Omega_I t}\cos(\Omega_R t)\,A_{\ell+2}
+\varepsilon\,c_\ell^{-}\,e^{\Omega_I t}\cos(\Omega_R t)\,A_{\ell-2}=0, 
\label{eq:chain}
\end{eqnarray}
where we have  assumed a   monochromatic Zerilli pump $H_i(t,r)=\Re[\hat H_i(r)e^{-i\Omega t}]$, 
with  $\Omega=\Omega_R+i\Omega_I$, $(\Omega_I<0)$. 

\section{ Two–mode reduction}

To proceed, let us truncate the GHZ–projected chain \eqref{eq:chain} to a parent-daughter pair $(\ell,\ell-2)$. The truncated amplitude  system is then written as 
\begin{align}
&\ddot A_\ell+\omega_\ell^2 A_\ell
+\varepsilon\,q_\ell\,e^{\Omega_I t}\cos(\Omega_R t)\,A_\ell
+\varepsilon\,c_\ell^{-}\,e^{\Omega_I t}\cos(\Omega_R t)\,A_{\ell-2}=0,
\label{eq:pair}\\
&\ddot A_{\ell-2}+\omega_{\ell-2}^2 A_{\ell-2}
+\varepsilon\,q_{\ell-2}\,e^{\Omega_I t}\cos(\Omega_R t)\,A_{\ell-2}
+\varepsilon\,c_{\ell-2}^{+}\,e^{\Omega_I t}\cos(\Omega_R t)\,A_{\ell}=0.
\nonumber
\end{align}
In order to derive the coupled-mode equations  and the growth rates near resonances from  the system of Eqs (\ref{eq:pair}), we will employ   the van der Pol–Krylov–Bogoliubov averaging method, where slowly varying envelopes are introduced and subsequently, an averaging over the fast phase is performed \cite{Nayfeh1973}.  Under the averaging (or rotating–wave approximation), only terms whose phases vary slowly survive,\footnote{In fact, one averages the equation over time windows $T$ such that $\omega_\ell T\gg 1$ while $|\Delta_\ell|\,T=\mathcal{O}(1)$. Rapidly oscillating pieces integrate to zero to leading order in $1/(\omega_\ell T)$.}
Therefore, we introduce the narrow–band complex envelopes $a_\ell(t)$ such that 
\begin{align}
A_\ell(t)=&\ a_\ell(t)\,e^{-i\omega_\ell^R t}+a_\ell^*(t)\,e^{+i\omega_\ell^R t},\nonumber \\
A_{\ell-2}(t)=&\ a_{\ell-2}(t)\,e^{-i\omega_{\ell-2}^R t}+a_{\ell-2}^*(t)\,e^{+i\omega_{\ell-2}^R t}. 
\label{eq:narrowband}
\end{align}
with the slow–variation conditions
\begin{equation}
|\dot a_\ell|\ll \omega_\ell^{R}|a_\ell|,\qquad |\ddot a_\ell|\ll \omega_\ell^{R}|\dot a_\ell|
,
\label{eq:env}
\end{equation}
and $\omega_\ell=\omega_\ell^R-i \alpha _\ell$ with $\alpha_\ell\ll\omega_\ell^R$. 
 Substituting (\ref{eq:narrowband}) into
(\ref{eq:pair}) and  keeping only near–stationary
phase terms,
we obtain, 
 after averaging over the fast carriers,
 the first–order envelope system
 \begin{align}
\dot a_\ell &=
-\alpha_\ell\,a_\ell
-\frac{i\varepsilon}{4\omega_\ell^{R}}\,q_\ell\,e^{\Omega_I t}\,a_\ell\,e^{+i\sigma_{\ell} t}
-\frac{i\varepsilon}{2\omega_\ell^{R}}\,c_\ell^{-}\,e^{\Omega_I t}\,a_{\ell-2}\,e^{+i\Delta_\ell t}, 
\label{eq:env-ell}\\
\dot a_{\ell-2} &=
-\alpha_{\ell-2}\,a_{\ell-2}
-\frac{i\varepsilon}{4\omega_{\ell-2}^{R}}\,q_{\ell-2}\,e^{\Omega_I t}\,a_{\ell-2}\,e^{+i\sigma_{\ell-2} t}
-\frac{i\varepsilon}{2\omega_{\ell-2}^{R}}\,c_{\ell-2}^{+}\,e^{\Omega_I t}\,a_{\ell}\,e^{-i\Delta_\ell t}.
\label{eq:env-ellm2}
\end{align}
where $\alpha_\ell=-\Im\,\omega_\ell>0$ and the  detunings are
\begin{eqnarray}
   \Delta_\ell=\Omega_R-(\omega_\ell^{R}-\omega_{\ell-2}^{R}),\qquad
\sigma_{\ell}=\Omega_R-2\omega_\ell^{R}.
\label{eq:detuning}
\end{eqnarray}
Equations \eqref{eq:env-ell}–\eqref{eq:env-ellm2} show explicitly the two
near–resonant channels: (i) the difference–frequency or 3-wave mixing  
($\Delta_\ell\approx 0$, off–diagonal terms) and
(ii) the diagonal Mathieu self–modulation ($\sigma_\ell\approx 0$, diagonal terms).
Their instability regions and growth rates  will be used determined below

 \subsection{ Off-diagonal, three-wave mixing channel }
\begin{figure}[H]
  \centering
  \includegraphics[width=0.65\linewidth]{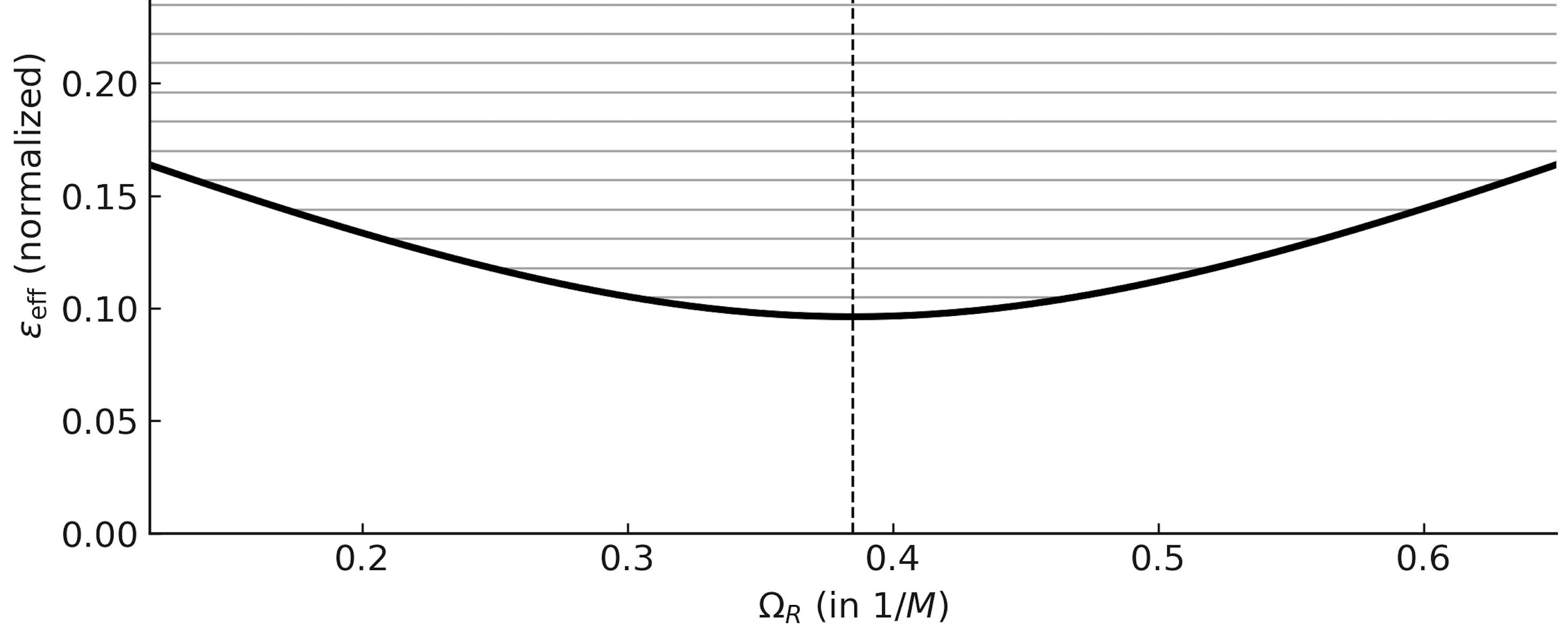}
  \caption{Arnold  instability tongue for the difference–frequency (three–wave) channel.
  The solid curve is the normalized threshold
  $\tilde{\varepsilon}_{\mathrm{thr}}(\Omega_R)=\sqrt{\alpha^{2}+\tfrac{1}{4}\big[\Omega_R-(\omega_\ell^R-\omega_{\ell-2}^R)\big]^2}$,
  and the horizontally hatched region denotes instability ($\tilde{\varepsilon}\!\ge\!\tilde{\varepsilon}_{\mathrm{thr}}$).
  The vertical dashed line marks the resonance center $\Omega_R=\omega_\ell^R-\omega_{\ell-2}^R$.
  For this plot we used the eikonal spacings $\omega_\ell^R-\omega_{\ell-2}^R=2\omega_c$ and $\alpha=\omega_c/2$, where $\omega_c=1/3\sqrt{3}M$. In practice, even low $\ell$ give the same Arnold tongues so the plot changes only by a small horizontal shift with a new minimum    and a vertical adjustment. For example, even $\ell=6$  scalar QNMs are already very close to eikonal values.  }
  \label{diff-tongue}
\end{figure}

 The first  near–resonant channel arises from the off–diagonal difference–frequency terms from $c_\ell^\pm$ (3-wave mixing).  In this channel only phases that are nearly stationary with detuning
\begin{equation}
\Delta_\ell=\Omega_R-(\omega_\ell^R-\omega_{\ell-2}^R)\approx 0.
\label{eq:Delta}
\end{equation}
survive the van der Pol–Krylov–Bogoliubov averaging.
Therfore, in this channel, the slow–envelope equations are
\begin{align}
\dot a_\ell &=
-\alpha_\ell\,a_\ell
-\frac{i\varepsilon}{2\omega_\ell^{R}}\,c_\ell^{-}\,e^{\Omega_I t}\,a_{\ell-2}\,e^{+i\Delta_\ell t}, 
\label{eq:env-ell1}\\
\dot a_{\ell-2} &=
-\alpha_{\ell-2}\,a_{\ell-2}
-\frac{i\varepsilon}{2\omega_{\ell-2}^{R}}\,c_{\ell-2}^{+}\,e^{\Omega_I t}\,a_{\ell}\,e^{-i\Delta_\ell t},
\label{eq:env-ell2}
\end{align}
where, the off–diagonal coefficients are
$c_\ell^{-}$ and $c_\ell^{-}$ are
\begin{align}
   c_\ell^{-}=\frac{G_{\ell,\ell-2}}{\mathcal N_\ell}\sum_{X\in\{tt,tr,rr,\theta\theta,\phi\phi\}}\!\!\mathscr{I}^{(X)}_{\ell,\ell-2},
\nonumber \\
c_{\ell-2}^{+}=\frac{G_{\ell-2,\ell}}{\mathcal N_{\ell-2}}\sum_{X\in\{tt,tr,rr,\theta\theta,\phi\phi\}}\mathscr{I}^{(X)}_{\ell-2,\ell},
\end{align}
where the angular factor $G_{\ell,\ell-2}$ is given by
\begin{eqnarray}
    G_{\ell,\ell-2}
=\frac{15}{2\pi}\,\frac{\ell^2(\ell-1)^2}{(2\ell)(2\ell-1)(2\ell-2)}, \label{eq:G}
\end{eqnarray}
$ \mathcal N_\ell$ is the GHZ norm $(\mathcal N_\ell=\l2 u_\ell,u_\ell\r2)$, and 
\begin{align}
\mathscr{I}^{(tt)}_{\ell\ell'}&=\l2 u_\ell,\ f^{-1}\hat H_0(-\omega_{\ell'}^2)u_{\ell'}+\tfrac12 f^{-1}(-i\Omega)(-i\omega_{\ell'})\hat H_0 u_{\ell'}\r2, \nonumber\\
\mathscr{I}^{(tr)}_{\ell\ell'}&=\l2 u_\ell,\ 2\hat H_1(i\omega_{\ell'})u'_{\ell'}+\Big((-i\Omega)\hat H_1+\tfrac{f'}{f}\hat H_1\Big)u'_{\ell'}\r2,\nonumber \\
\mathscr{I}^{(rr)}_{\ell\ell'}&=\l2 u_\ell,\ f\hat H_2 u''_{\ell'}+\Big(\tfrac{f'}{2}\hat H_2+\tfrac{f}{2}\hat H_2'\Big)u'_{\ell'}\r2,
\\
\mathscr{I}^{(\theta\theta)}_{\ell\ell'}&=\l2 u_\ell,\ \frac{\hat K}{r^2}\ell'(\ell'+1)u_{\ell'}+\frac{\hat K'}{r}u'_{\ell'}+\frac{(-i\Omega)\hat K}{fr}(i\omega_{\ell'})u_{\ell'}\r2, \nonumber \\
\mathscr{I}^{(\theta\theta)}_{\ell\ell'}&=\mathscr{I}^{(\phi\phi)}_{\ell\ell'}.
\label{eq:kernel}
\end{align}
By parametrizing $a_\ell$ and $a_{\ell-2}$ by new 
 narrow envelops $b_\ell$ and $b_{\ell-2}$ defined by 
\begin{equation}
a_\ell=b_\ell=\rho^{-1} e^{i\theta}\, b_\ell\,e^{i\Delta_\ell t/2},\qquad
a_{\ell-2}=\rho e^{-i\theta\, } b_{\ell-2}e^{-i\Delta_\ell t/2},
\label{eq:bell}
\end{equation}
where
\begin{eqnarray}
\rho=\sqrt{\frac{\omega_\ell\,|c_{\ell-2}^+|}{\omega_{\ell-2}\,|c_{\ell}^+|}},\qquad
    \theta=\frac{1}{2}\left(\pi-\arg c_{\ell-2}^+-\arg c_\ell^-\right),
\end{eqnarray} 
we find that Eqs (\ref{eq:env-ell1}) and (\ref{eq:env-ell2})  are written as  the linear $2\times2$ system 
\begin{equation}
\begin{pmatrix}
\dot b_\ell\\[3pt] \dot b_{\ell-2}
\end{pmatrix}
=
\underbrace{\begin{pmatrix}
-\alpha_\ell - i\frac{\Delta_\ell}{2} & i\,g(t)\\[3pt]
-i\,g(t) & -\alpha_{\ell-2} + i\frac{\Delta_\ell}{2}
\end{pmatrix}}_{= \ \mathcal{M}_{\rm d}(t)}
\begin{pmatrix}
b_\ell\\[3pt] b_{\ell-2}
\end{pmatrix}.
\label{eq:2x2-time}
\end{equation}
with 
\begin{eqnarray}
    g(t)=g_0e^{\Omega_I t}, \qquad 
    g_0=\frac{\varepsilon}{2}\sqrt{\frac{|c_\ell^-c_{\ell-2}^+}{\omega_{\ell}^R\omega_{\ell-2}^R}},
\end{eqnarray}
If $|\Omega_I|$ is small compared to all oscillation or damping rates, the growth can be determined adiabatically by treating $g(t)$ as frozen. The instantaneous Floquet exponents are the eigenvalues of $\mathcal{M}_{\rm d}(t)$,
\begin{equation}
\lambda_{\pm}(t)= -\frac{\alpha_\ell+\alpha_{\ell-2}}{2}
\pm
\frac{1}{2}\sqrt{\,4 g(t)^2+(a_\ell-a_{\ell-2})^2-\Delta_\ell^2 +2 i (a_\ell-a_{\ell-2})\Delta_\ell\ } \ .
\label{eq:eigs}
\end{equation}
Instability occurs whenever the eigenvalues have positive real part, i.e., 
\begin{eqnarray}
    {\rm Re}(\lambda_+)>0. 
\end{eqnarray}$(\alpha_\ell+\alpha_{\ell-2})/2$, 
where 
\begin{align}
    {\rm Re}(\lambda_+)=&-\frac{\alpha_\ell+\alpha_{\ell-2}}{2}+\frac{1}{2\sqrt{2}}\Bigg\{
    \bigg[\Big(4 g(t)^2+(a_\ell-a_{\ell-2})^2-\Delta_\ell^2\Big)^2+
    4  (a_\ell-a_{\ell-2})^2\Delta_\ell^2\bigg]^{1/2}
   \nonumber \\
   &+4 g(t)^2+(a_\ell-a_{\ell-2})^2-\Delta_\ell^2
    \Bigg\}^{1/2}
\end{align}
For balanced damping
with 
$\alpha_\ell=\alpha_{\ell-2}=\alpha$, we find that 
instability sets in ($\Re\,\lambda_+=0$) for the threshold value of $g$ 
\begin{equation}
g_{\mathrm{thr}}=\sqrt{\alpha^2+\frac{\Delta_\ell^2}{4}}.
\end{equation}
Expressing $g$ through the physical modulation depth $\varepsilon$ gives
\begin{equation}
\varepsilon_{\mathrm{thr}}(\Omega_R,t)
= \,e^{-\Omega_I t}\,\sqrt{\frac{\omega_\ell^R\omega_{\ell-2}^R}{|c_\ell^- c_{\ell-2}^+|}}\,
\sqrt{4\alpha^2+\Big[\Omega_R-(\omega_\ell^R-\omega_{\ell-2}^R)\Big]^2}.
\end{equation}
It is convenient to factor out the unknown coupling normalization and define the normalized threshold
\begin{equation}
\widetilde{\varepsilon}_{\mathrm{thr}}(\Omega_R)
= \frac{\varepsilon_{\mathrm{thr}}}{2}\,
\sqrt{\frac{|c_\ell^- c_{\ell-2}^+|}{2\omega_\ell^R\omega_{\ell-2}^R}}\ e^{\Omega_I t}
= \frac{1}{2}\sqrt{4\alpha^2+\Big[\Omega_R-(\omega_\ell^R-\omega_{\ell-2}^R)\Big]^2}.
\end{equation}
The difference–frequency resonance tongue for large $\ell$ is plotted in Fig. \ref{diff-tongue}. 

\noindent
We note also that three-wave mixing appears in rotating  stars arising from quadratic mode couplings. In this case,  quadratic perturbations couple the unstable $f$‐mode to otherwise stable daughter modes and when near–resonance conditions are met,  three–wave mixing  triggers parametric resonance, redistributing energy among the modes and saturating the $f$‐mode amplitude \cite{Pnigouras:2015bwa,Pnigouras:2016gun}.


\subsection{Diagonal Mathieu channel}
 
\begin{figure}[H]
  \centering
  \includegraphics[width=0.55\linewidth]{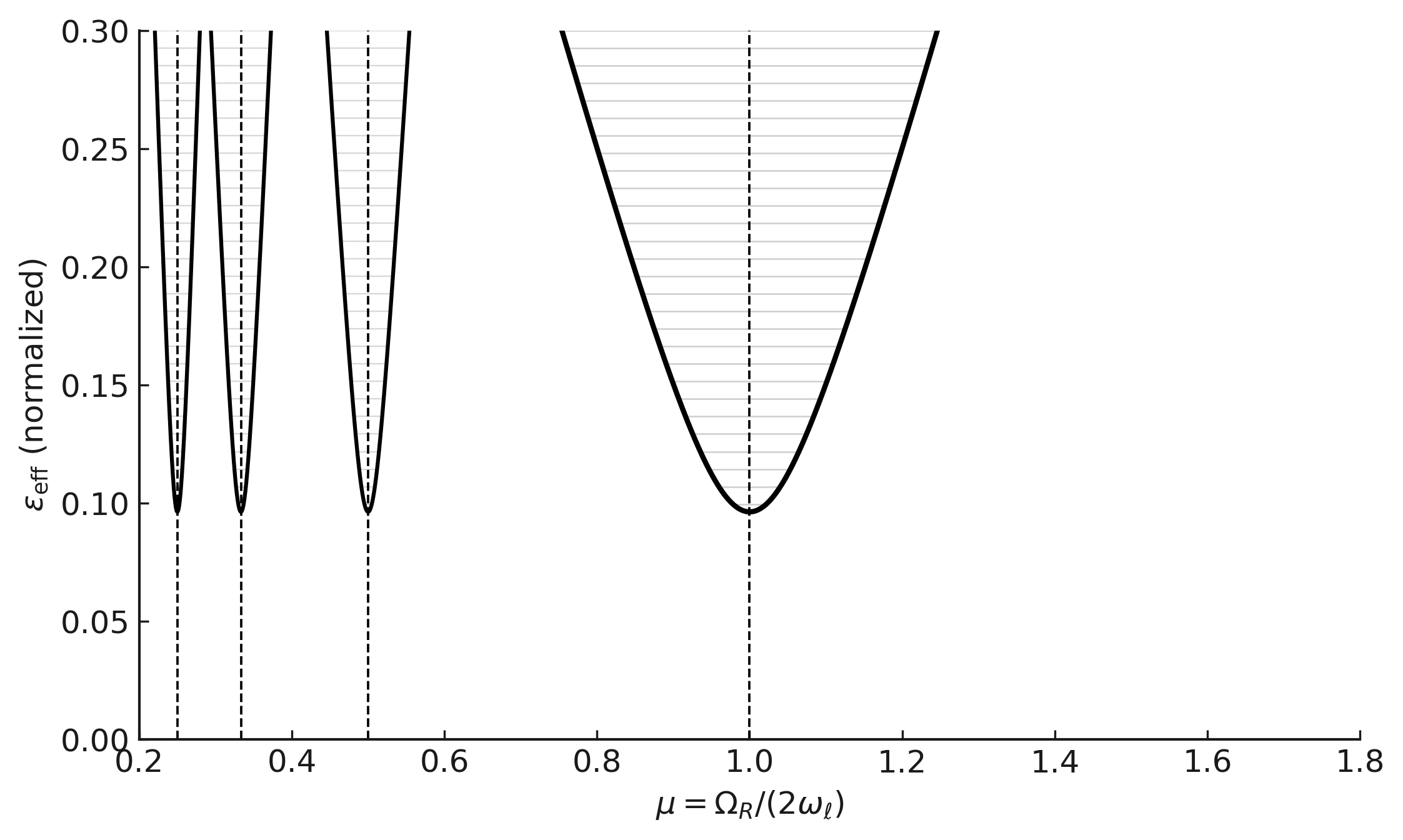}
  \caption{Mathieu–type parametric–instability map  in the scaled plane
$\mu=\Omega_R/(2\omega_\ell)$ and $\varepsilon_{\mathrm{eff}}=\varepsilon |q_\ell|/(4\omega_\ell)$.
Black curves are the instability boundaries; horizontal grey lines indicate the unstable
region. The primary lobe is centered at $\mu=1$; higher–order
tongues at $\mu=\tfrac12,\tfrac13,\tfrac14$ are shown narrower, as expected from
higher–order averaging. Vertical dashed lines mark the tongue centers.}
  \label{mathieu-tongue}
\end{figure}

The second near–resonant instability arises from the  diagonal Mathieu, self-parametric channel. 
In this case, the diagonal modulation within the narrow–band  approximation of Eqs (\ref{eq:env-ell}) and (\ref{eq:env-ellm2}) turns out to be
\begin{align}
    \dot a_\ell=&-\alpha_\ell a_\ell-\frac{i\varepsilon}{4\omega_\ell^{R}}\,q_\ell\,e^{\Omega_I t}\,a_\ell^*\,e^{+i\sigma_{\ell} t},
\nonumber \\
\dot a_\ell^*=&-\alpha_\ell a_\ell^*+\frac{i\varepsilon}{4\omega_\ell^{R}}\,q_\ell^*\,e^{\Omega_I t}\,a_\ell\,e^{-i\sigma_{\ell} t}, \qquad \sigma_\ell=2\omega_\ell-\Omega_R\simeq 0.
\label{eq:system}
\end{align}
We may now define new envelopes $b_\ell$ and $b_\ell^*$ such that
\begin{eqnarray}
    a_\ell=b_\ell e^{+i \sigma_\ell t/2}, \qquad a_\ell^*=b_\ell^* e^{-i \sigma_\ell t/2}
\end{eqnarray}
and let  $p(t)=\frac{\varepsilon|q_\ell|}{4\omega_\ell}e^{\Omega_I t}$.
Then,   Eq. (\ref{eq:system}) is written as  
\begin{eqnarray}
    \begin{pmatrix} \dot b_\ell\\ \dot 
 b_\ell^*\end{pmatrix}
=
\underbrace{\begin{pmatrix}
-\alpha_\ell - i\sigma_\ell/2 & \ \ -i\, p(t)e^{i\phi_\ell}\\[2pt]
\ \ +i \,p(t)e^{-i\phi_\ell} & -\alpha_\ell + i\sigma_\ell/2
\end{pmatrix}}_{\mathcal{M}_\sigma(t)}
\!\begin{pmatrix} b_\ell\\ b^*_\ell\end{pmatrix},
\label{eq:Mm}
\end{eqnarray}
where $\phi=\arg q_\ell$. The eigenvalues of the matrix $\mathcal{M}_\sigma(t)$ are 
\begin{eqnarray}
    \lambda_\pm=-\alpha_\ell\pm\frac{1}{2} \sqrt{4 p(t)^2-\sigma_\ell^2}.
\end{eqnarray}
Then, instability occurs whenever ${\rm Re}(\lambda_+)>0$, which is the case for 
\begin{eqnarray}
p(t)>\frac{1}{2}\sqrt{4\alpha_\ell^2+\sigma_\ell^2}. 
\end{eqnarray}
In other words, instability is set in for the threshold value of $p(t)$
\begin{eqnarray}
p_{\rm th}(t)=\frac{1}{2}\sqrt{4\alpha_\ell^2+\big(2\omega^R_\ell-\Omega_R\big)^2}, 
\end{eqnarray}
which corresponds to the threshold for  physical modulation depth $\epsilon$ 
\begin{eqnarray}
\varepsilon_{\rm th}(\Omega,t)=e^{-\Omega_I t}\frac{2\omega^R_\ell}{|q_\ell|}\sqrt{4\alpha_\ell^2+\big(2\omega^R_\ell-\Omega_R\big)^2}. 
\end{eqnarray}
Again, it is convenient to factor out the unknown coupling normalization and define the normalized threshold
\begin{eqnarray}
\widetilde\varepsilon_{\rm th}(\Omega_R,\omega^R_\ell)=\frac{\varepsilon_{\rm th} |q_\ell|}{4\omega_\ell}e^{\Omega_I t}=\frac{1}{2}\sqrt{4\alpha_\ell^2+\big(2\omega_\ell^R-\Omega_R\big)^2}. 
\end{eqnarray}
Therefore the Mathieu instability  is centered at $\Omega_R=2\omega^R_\ell$ with boundary
$\widetilde\varepsilon_{\rm th}$. 
At exact tuning $(\sigma_\ell=0)$ one has  $\widetilde\varepsilon_{\rm th}=\alpha_\ell$ (e.g.\ $p(0)>\alpha_\ell$ for a decaying pump $\Omega_I<0$) to obtain net amplification within the pump lifetime.
The Arnold tongue for the Mathieu channel centered at $\mu=1$ is plotted in Fig. (\ref{mathieu-tongue}). One may notice  that there are additional tongues centered in Fig. (\ref{mathieu-tongue}) for $\mu=\tfrac12,\tfrac13\cdots$, which we will explain below.

We should stress here that the instabilities we  described above is in fact a transient instability. The reason is that the matrices $\mathcal{M}_{\rm d}(t)$ and  $\mathcal{M}_\sigma(t)$ are functions of time. For the particular $\mathcal{M}_\sigma(t)$ for example, the positivity of the maximum eigenvalue, although it implies transient instability, it does not imply global time instability. In fact, the solution of the system of Eq. (\ref{eq:Mm}) 
is bounded by 
\begin{eqnarray}
    b_\ell(t)<b_\ell(0) e^{\int_0^t(-\alpha_\ell-|p(\tau)|)\d \tau}.
\end{eqnarray}
Then, it is easy to verify that transient growth exists up to a time scale 
\begin{eqnarray}
    t_{\rm s}=\frac{1}{|\Omega_I|} \ln\left(\frac{p(0)}{\alpha_\ell}\right),
\end{eqnarray}
after which we end up in exponential decay. 
In other words, there is no long-time parametric divergence with a decaying pump ($e^{\Omega_I t})$ since only bounded transient growth is possible.

\section{Higher-order instability}
These tongues are originated from  nonlinear GR as we will explain below. 
In our setup the scalar amplitude $A_\ell$ experiences a parametric modulation because
the background metric is not strictly stationary: it is weakly time–periodic with frequency $\Omega_R$
due to an $L=2,M=0$ pump. At first order in the metric amplitude (or $\varepsilon$), we
found a single harmonic proportional to $ \cos\Omega_R t$ in the coefficients of the scalar master equation. 
At higher orders of the Einstein perturbation expansion, new harmonics $n\Omega_R$ necessarily appear,
and these are precisely the harmonics that generate the tongues centered at $\mu=\Omega_R/(2\omega^R_\ell)=1/n$.

In order to find out here the higher harmonics $n\Omega_R$ come from, 
let us recall that the metric is expanded as
\begin{equation}
g_{ab}=g^{(0)}_{ab}+\varepsilon\,h^{(1)}_{ab}+\varepsilon^2 h^{(2)}_{ab}+\varepsilon^3 h^{(3)}_{ab}+\cdots,
\label{eq:metric-expansion}
\end{equation}
with $g^{(0)}_{ab}$ the backgroudn Schwarzschild  metric and $h^{(1)}_{ab}(t,\mathbf{x})=\Re\!\big[\hat H^{(1)}_{ab}(\mathbf{x})\,e^{-i\Omega_R t}\big]$ the first–order pump  with $L=2,M=0$.
Einstein’s equations $E[g]=0$ expand schematically as
\begin{align}
&\mathcal{E}[h^{(1)}] \;=\;0,
\label{eq:Ein1}
\\
&\mathcal{E}[h^{(2)}] \;=\; S^{(2)}[h^{(1)},h^{(1)}],
\qquad
\mathcal{E}[h^{(3)}] \;=\; S^{(3)}[h^{(1)},h^{(2)}]+\tilde S^{(3)}[h^{(1)},h^{(1)},h^{(1)}],
\quad \text{etc.}
\label{eq:Ein23}
\end{align}
Here $\mathcal{E}$ is the linearized Einstein operator about Schwarzschild,
and $S^{(k)}$ are multilinear source functionals built from products of lower–order fields and their derivatives.
Because $h^{(1)}\! \sim \! e^{-i\Omega_R t}$, the quadratic source contains harmonics
\begin{eqnarray}
   S^{(2)}[h^{(1)},h^{(1)}]\;\sim\;
\{\big(e^{-i\Omega_R t}\big)^2,\; e^{-i\Omega_R t}e^{+i\Omega_R t},\;\big(e^{+i\Omega_R t}\big)^2\}
=
\{\,e^{-2i\Omega_R t},\; 1,\; e^{+2i\Omega_R t}\,\}. 
\end{eqnarray}
Thus, the second–order metric $h^{(2)}$ contains a component oscillating at $2\Omega_R$, as well as  a constant component.
At third order one similarly obtains $h^{(3)}$ with harmonics at $3\Omega_R$, and so on. Therefore the
time–dependence of the effective coefficients in the scalar master equation inevitably involves a Fourier
series, since Eq. (\ref{eq:chain}) has the schematic form, up to $k$-th order, 
\begin{eqnarray}
&&\ddot A_\ell+\omega_\ell^2 A_\ell
+\sum_{n=1}^k\ \hspace{1mm}
\sum_{|l-2n|\leq l\leq l+2n}  
\varepsilon^n \ C_{mn}^{(\ell)} \,e^{n\,\Omega_I t}\cos(n\,\Omega_R t)\,A_m
=0.
\label{eq:chain11}
\end{eqnarray} 
The appearance of $2n+1$ terms in Eq. (\ref{eq:chain11}) is due to the 
spherical harmonic relation 
\begin{eqnarray}
\underbrace{Y_{20}\,Y_{20}\cdots Y_{20}}_{n\ \text{factors}}
\;=\;\sum_{L\in\{0,2,4,\dots,2n\}} c^{(n)}_{L}\,Y_{L0}. 
\end{eqnarray}
Now, each harmonic $\cos(n\,\Omega_R t)$ acts as an
$n$-th harmonic parametric driver. Consequently, in the self–modulation, $A_\ell$ satisfies 
\begin{eqnarray}
&&\ddot A_\ell+\omega_\ell^2 A_\ell
+\Bigg[\sum_{n=1}^k
\varepsilon^n \ C_{\ell n}^{(\ell)} \,e^{n\,\Omega_I t}\cos(n\,\Omega_R t)\Bigg]A_\ell
=0.
\label{eq:chain1}
\end{eqnarray} 
so that
the near–resonant phase that survives averaging is
$e^{i(2\omega^R_\ell-n\Omega_R)t}$, producing the standard Mathieu resonance condition
\begin{equation}
2\omega^R_\ell \approx n\,\Omega_R
\qquad\Longleftrightarrow\qquad \mu=\frac{\Omega_R}{2\omega^R_\ell}\approx\frac{1}{n},
\qquad n=1,2,3,\ldots \  \ .
\label{eq:centers-nGR}
\end{equation}
Hence, there are now Arnold tongues centered at $\mu=1/n$, which are the ones depicted in Fig. (\ref{mathieu-tongue}).  Importantly, the order in perturbation theory that first
produces the $n$-th harmonic also sets the scaling of its strength. The amplitude of the $n$th
harmonic is generically ${\cal O}(\varepsilon^{\,n})$ in GR perturbation theory, which yields the 
shrinking of the higher tongues  we see in Fig. \ref{mathieu-tongue}.

The discussion above concerns the diagonal, self–modulation channel, where $a_\ell$ couples to its complex
conjugate $a_\ell^*$. In contrast, the difference–frequency (three–wave) channel couples $a_\ell$ and
$a_{\ell-2}$ with a near–stationary phase $e^{i(\omega_\ell-\omega_{\ell-2}-\Omega_R)t}$, producing a single strong
tongue centered at $\Omega_R\simeq \omega^R_\ell-\omega^R_{\ell-2}$ (which tends to $2\Omega_c$ in the eikonal limit). 
Additional comb–like tongues do not arise  since for the $n$th harmonic we will have 
\begin{eqnarray}
    \Omega_R=\frac{\omega_\ell^R-\omega_{\ell-2}^R}{n}. \label{eq:omega}
\end{eqnarray}
Since, $\omega_\ell^R-\omega_{\ell-2}^R\approx 2 \omega_c$, there is  a single-tone pump $\Omega_R$ that satisfies (\ref{eq:omega}) for a given $n$ (taken to be the leading $n=1$), so that $\Omega_R\approx2\omega_c$. Therefore, no higher Arnold tongues appear in the three-wave chanel.

A final comment concerns  $\varepsilon$, which although it has been introduced  as a bookkeeping parameter in first place, it has acquired physical meaning as the dimensionless pump amplitude. This is possible since
if one rescales the pump as $\hat H_{ab}\!\to\!A\,\hat H_{ab}$ and simultaneously sets
$\varepsilon\!\to\!1$, all formulas are unchanged except that $\varepsilon$ is replaced by
the physical amplitude $A$. 
In practice, we either  keep $\varepsilon$ explicit and calculate
$\varepsilon_{\rm thr} $,
or set $\varepsilon=1$ and state the
threshold directly as a condition on the pump amplitude (or on the norm of $\hat H_{ab}$)
as $A_{\rm thr}$
We have chosen the first option. 

\section{Conclusions}
Nonlinear interactions of gravity generate a plethora of unexpected phenomena, rendering the detection   of BH QNMs  an unique occasion to learn about the properties of gravity. 
The nonlinear dynamics explored here shed also light on binary BH behavior. During merger, equal-mass circular systems primarily excite the $\ell=2$ mode, with frequency increasing during inspiral. Our results, as well as the  recent ones in the literature \cite{Ma:2025rnv}, seem to indicate that  resonant interactions tend to  preserve the lowest angular and frequency modes. This has  important consequences from the observational side, e.g. on the perspective to detect the quadratic QNMs. Of course our results are based on the impact on a scalar probe, the next necessary step will be to generalize our findings to the gravitational degrees of freedom.

\normalsize
\begin{acknowledgments}
\noindent
We would like to thank K. Kokkotas for illuminating discussions.      A.R.  acknowledges support from the  Swiss National Science Foundation (project number CRSII5\_213497).
\end{acknowledgments}

\bibliographystyle{JHEP}
\bibliography{draft}

@article{Kehagias:2025ntm,
    author = "Kehagias, Alex and Perrone, Davide and Riotto, Antonio",
    title = "{Non-linear Quasi-Normal Modes of the Schwarzschild Black Hole from the Penrose Limit}",
    eprint = "2503.09350",
    archivePrefix = "arXiv",
    primaryClass = "gr-qc",
    month = "3",
    year = "2025"
}

@article{Kehagias:2025gvk,
    author = "Kehagias, Alex and Perrone, Davide and Riotto, Antonio",
    title = "{Nonlinearities of Schwarzschild Black Hole Head-on Collisions}",
    eprint = "2508.17993",
    archivePrefix = "arXiv",
    primaryClass = "gr-qc",
    month = "8",
    year = "2025"
}

@article{Ianniccari:2025avm,
    author = "Ianniccari, A. and Lo Bianco, L. and Riotto, A.",
    title = "{The nonlinear tails in black hole ringdown: the scattering perspective}",
    eprint = "2507.17732",
    archivePrefix = "arXiv",
    primaryClass = "gr-qc",
    doi = "10.1088/1475-7516/2025/10/062",
    journal = "JCAP",
    volume = "10",
    pages = "062",
    year = "2025"
}

@article{Kehagias:2025xzm,
    author = "Kehagias, Alex and Riotto, Antonio",
    title = "{Nonlinear Tails of Gravitational Waves in Schwarzschild Black Hole Ringdown}",
    eprint = "2504.06224",
    archivePrefix = "arXiv",
    primaryClass = "gr-qc",
    month = "4",
    year = "2025"
}

@article{Ling:2025wfv,
    author = "Ling, Siyang and Shah, Sabeela and Wong, Sam S. C.",
    title = "{Dynamical nonlinear tails in Schwarzschild black hole ringdown}",
    eprint = "2503.19967",
    archivePrefix = "arXiv",
    primaryClass = "gr-qc",
    month = "3",
    year = "2025"
}

@article{Kehagias:2025tqi,
    author = "Kehagias, Alex and Riotto, Antonio",
    title = "{The AdS Perspective on the Nonlinear Tails in Black Hole Ringdown}",
    eprint = "2506.14475",
    archivePrefix = "arXiv",
    primaryClass = "gr-qc",
    month = "6",
    year = "2025"
}

@misc{shi2024detectabilityresolvabilityquasinormalmodes,
      title={On the detectability and resolvability of quasi-normal modes with space-based gravitational wave detectors}, 
      author={Changfu Shi and Qingfei Zhang and Jianwei Mei},
      year={2024},
      eprint={2407.13110},
      archivePrefix={arXiv},
      primaryClass={gr-qc},
      url={https://arxiv.org/abs/2407.13110}, 
}

@article{Yi_2024,
   title={Nonlinear quasinormal mode detectability with next-generation gravitational wave detectors},
   volume={109},
   ISSN={2470-0029},
   url={http://dx.doi.org/10.1103/PhysRevD.109.124029},
   DOI={10.1103/physrevd.109.124029},
   number={12},
   journal={Physical Review D},
   publisher={American Physical Society (APS)},
   author={Yi, Sophia and Kuntz, Adrien and Barausse, Enrico and Berti, Emanuele and Cheung, Mark Ho-Yeuk and Kritos, Konstantinos and Maselli, Andrea},
   year={2024},
   month=jun }

@article{PhysRevD.109.064075,
  title = {Linear versus nonlinear modeling of black hole ringdowns},
  author = {Qiu, Yi and Forteza, Xisco Jim\'enez and Mourier, Pierre},
  journal = {Phys. Rev. D},
  volume = {109},
  issue = {6},
  pages = {064075},
  numpages = {18},
  year = {2024},
  month = {Mar},
  publisher = {American Physical Society},
  doi = {10.1103/PhysRevD.109.064075},
  url = {https://link.aps.org/doi/10.1103/PhysRevD.109.064075}
}

@article{Berti:2025hly,
    author = "Berti, Emanuele and others",
    title = "{Black hole spectroscopy: from theory to experiment}",
    eprint = "2505.23895",
    archivePrefix = "arXiv",
    primaryClass = "gr-qc",
    month = "5",
    year = "2025"
}

@article{Singh:2025xzd,
    author = "Singh, Jasveer and Suneeta, Vardarajan",
    title = "{Computing nonlinearity ratios using second order black hole perturbation theory}",
    eprint = "2512.00943",
    archivePrefix = "arXiv",
    primaryClass = "gr-qc",
    month = "11",
    year = "2025"
}

@article{Bucciotti:2025rxa,
    author = "Bucciotti, Bruno and Cardoso, Vitor and Kuntz, Adrien and Pere\~niguez, David and Redondo-Yuste, Jaime",
    title = "{Ringdown nonlinearities in the eikonal regime}",
    eprint = "2501.17950",
    archivePrefix = "arXiv",
    primaryClass = "gr-qc",
    month = "1",
    year = "2025"
}

@article{Lagos:2024ekd,
    author = "Lagos, Macarena and Andrade, Tom\'as and Rafecas-Ventosa, Jordi and Hui, Lam",
    title = "{Black hole spectroscopy with nonlinear quasinormal modes}",
    eprint = "2411.02264",
    archivePrefix = "arXiv",
    primaryClass = "gr-qc",
    doi = "10.1103/PhysRevD.111.024018",
    journal = "Phys. Rev. D",
    volume = "111",
    number = "2",
    pages = "024018",
    year = "2025"
}

@article{LIGOScientific:2021sio,
    author = "Abbott, R. and others",
    collaboration = "LIGO Scientific, VIRGO, KAGRA",
    title = "{Tests of General Relativity with GWTC-3}",
    eprint = "2112.06861",
    archivePrefix = "arXiv",
    primaryClass = "gr-qc",
    reportNumber = "LIGO-P2100275",
    month = "12",
    year = "2021"
}

@article{LIGOScientific:2020tif,
    author = "Abbott, R. and others",
    collaboration = "LIGO Scientific, Virgo",
    title = "{Tests of general relativity with binary black holes from the second LIGO-Virgo gravitational-wave transient catalog}",
    eprint = "2010.14529",
    archivePrefix = "arXiv",
    primaryClass = "gr-qc",
    reportNumber = "LIGO-P2000091",
    doi = "10.1103/PhysRevD.103.122002",
    journal = "Phys. Rev. D",
    volume = "103",
    number = "12",
    pages = "122002",
    year = "2021"
}

@article{London:2014cma,
    author = "London, Lionel and Shoemaker, Deirdre and Healy, James",
    title = "{Modeling ringdown: Beyond the fundamental quasinormal modes}",
    eprint = "1404.3197",
    archivePrefix = "arXiv",
    primaryClass = "gr-qc",
    doi = "10.1103/PhysRevD.90.124032",
    journal = "Phys. Rev. D",
    volume = "90",
    number = "12",
    pages = "124032",
    year = "2014",
    note = "[Erratum: Phys.Rev.D 94, 069902 (2016)]"
}

@article{Khera:2024yrk,
    author = "Khera, Neev and Ma, Sizheng and Yang, Huan",
    title = "{Quadratic Mode Couplings in Rotating Black Holes and Their Detectability}",
    eprint = "2410.14529",
    archivePrefix = "arXiv",
    primaryClass = "gr-qc",
    month = "10",
    year = "2024"
}

@article{Cheung:2022rbm,
    author = "Cheung, Mark Ho-Yeuk and others",
    title = "{Nonlinear Effects in Black Hole Ringdown}",
    eprint = "2208.07374",
    archivePrefix = "arXiv",
    primaryClass = "gr-qc",
    doi = "10.1103/PhysRevLett.130.081401",
    journal = "Phys. Rev. Lett.",
    volume = "130",
    number = "8",
    pages = "081401",
    year = "2023"
}

@article{Pnigouras:2015bwa,
    author = "Pnigouras, Pantelis and Kokkotas, Kostas D.",
    title = "{Saturation of the f -mode instability in neutron stars: Theoretical framework}",
    eprint = "1509.01453",
    archivePrefix = "arXiv",
    primaryClass = "astro-ph.HE",
    doi = "10.1103/PhysRevD.92.084018",
    journal = "Phys. Rev. D",
    volume = "92",
    number = "8",
    pages = "084018",
    year = "2015"
}

@article{Pnigouras:2016gun,
    author = "Pnigouras, Pantelis and Kokkotas, Kostas D.",
    title = "{Saturation of the f-mode instability in neutron stars: II. Applications and results}",
    eprint = "1607.03059",
    archivePrefix = "arXiv",
    primaryClass = "astro-ph.HE",
    doi = "10.1103/PhysRevD.94.024053",
    journal = "Phys. Rev. D",
    volume = "94",
    number = "2",
    pages = "024053",
    year = "2016"
}

@article{Pitte:2024zbi,
    author = "Pitte, Chantal and Baghi, Quentin and Besan\c{c}on, Marc and Petiteau, Antoine",
    title = "{Exploring tests of the no-hair theorem with LISA}",
    eprint = "2406.14552",
    archivePrefix = "arXiv",
    primaryClass = "gr-qc",
    doi = "10.1103/PhysRevD.110.104003",
    journal = "Phys. Rev. D",
    volume = "110",
    number = "10",
    pages = "104003",
    year = "2024"
}

@article{Bucciotti:2024zyp,
    author = "Bucciotti, Bruno and Juliano, Leonardo and Kuntz, Adrien and Trincherini, Enrico",
    title = "{Quadratic Quasi-Normal Modes of a Schwarzschild Black Hole}",
    eprint = "2405.06012",
    archivePrefix = "arXiv",
    primaryClass = "gr-qc",
    month = "5",
    year = "2024"
}

@article{Zhu:2024rej,
    author = "Zhu, Hengrui and others",
    title = "{Nonlinear effects in black hole ringdown from scattering experiments: Spin and initial data dependence of quadratic mode coupling}",
    eprint = "2401.00805",
    archivePrefix = "arXiv",
    primaryClass = "gr-qc",
    doi = "10.1103/PhysRevD.109.104050",
    journal = "Phys. Rev. D",
    volume = "109",
    number = "10",
    pages = "104050",
    year = "2024"
}

@article{Redondo-Yuste:2023seq,
    author = "Redondo-Yuste, Jaime and Carullo, Gregorio and Ripley, Justin L. and Berti, Emanuele and Cardoso, Vitor",
    title = "{Spin dependence of black hole ringdown nonlinearities}",
    eprint = "2308.14796",
    archivePrefix = "arXiv",
    primaryClass = "gr-qc",
    doi = "10.1103/PhysRevD.109.L101503",
    journal = "Phys. Rev. D",
    volume = "109",
    number = "10",
    pages = "L101503",
    year = "2024"
}

@article{Kehagias:2023ctr,
    author = "Kehagias, Alex and Perrone, Davide and Riotto, Antonio and Riva, Francesco",
    title = "{Explaining nonlinearities in black hole ringdowns from symmetries}",
    eprint = "2301.09345",
    archivePrefix = "arXiv",
    primaryClass = "gr-qc",
    doi = "10.1103/PhysRevD.108.L021501",
    journal = "Phys. Rev. D",
    volume = "108",
    number = "2",
    pages = "L021501",
    year = "2023"
}

@article{Finch:2022ynt,
    author = "Finch, Eliot and Moore, Christopher J.",
    title = "{Searching for a ringdown overtone in GW150914}",
    eprint = "2205.07809",
    archivePrefix = "arXiv",
    primaryClass = "gr-qc",
    reportNumber = "LIGO document number P2200149",
    doi = "10.1103/PhysRevD.106.043005",
    journal = "Phys. Rev. D",
    volume = "106",
    number = "4",
    pages = "043005",
    year = "2022"
}

@article{Capano:2021etf,
    author = "Capano, Collin D. and Cabero, Miriam and Westerweck, Julian and Abedi, Jahed and Kastha, Shilpa and Nitz, Alexander H. and Wang, Yi-Fan and Nielsen, Alex B. and Krishnan, Badri",
    title = "{Multimode Quasinormal Spectrum from a Perturbed Black Hole}",
    eprint = "2105.05238",
    archivePrefix = "arXiv",
    primaryClass = "gr-qc",
    doi = "10.1103/PhysRevLett.131.221402",
    journal = "Phys. Rev. Lett.",
    volume = "131",
    number = "22",
    pages = "221402",
    year = "2023"
}

@article{Kehagias:2024sgh,
    author = "Kehagias, A. and Riotto, A.",
    title = "{Nonlinear effects in black hole ringdown made simple: Quasinormal modes as adiabatic modes}",
    eprint = "2411.07980",
    archivePrefix = "arXiv",
    primaryClass = "gr-qc",
    doi = "10.1103/PhysRevD.111.L041506",
    journal = "Phys. Rev. D",
    volume = "111",
    number = "4",
    pages = "L041506",
    year = "2025"
}

@article{Perrone:2023jzq,
    author = "Perrone, Davide and Barreira, Thomas and Kehagias, Alex and Riotto, Antonio",
    title = "{Non-linear black hole ringdowns: An analytical approach}",
    eprint = "2308.15886",
    archivePrefix = "arXiv",
    primaryClass = "gr-qc",
    doi = "10.1016/j.nuclphysb.2023.116432",
    journal = "Nucl. Phys. B",
    volume = "999",
    pages = "116432",
    year = "2024"
}

@article{Perrone:2025zhy,
    author = "Perrone, Davide and Kehagias, Alex and Riotto, Antonio",
    title = "{Nonlinearities in Kerr black hole ringdown from the Penrose limit}",
    eprint = "2507.01919",
    archivePrefix = "arXiv",
    primaryClass = "gr-qc",
    doi = "10.1088/1475-7516/2025/10/024",
    journal = "JCAP",
    volume = "10",
    pages = "024",
    year = "2025"
}

@article{Figueras:2023ihz,
    author = "Figueras, Pau and Rossi, Lorenzo",
    title = "{Non-linear instability of slowly rotating Kerr-AdS black holes}",
    eprint = "2311.14167",
    archivePrefix = "arXiv",
    primaryClass = "hep-th",
    doi = "10.1007/JHEP06(2025)107",
    journal = "JHEP",
    volume = "06",
    pages = "107",
    year = "2025"
}

@article{Galtier:2017mve,
    author = "Galtier, S{\'e}bastien and Nazarenko, Sergey V.",
    title = "{Turbulence of Weak Gravitational Waves in the Early Universe}",
    eprint = "1703.09069",
    archivePrefix = "arXiv",
    primaryClass = "gr-qc",
    doi = "10.1103/PhysRevLett.119.221101",
    journal = "Phys. Rev. Lett.",
    volume = "119",
    number = "22",
    pages = "221101",
    year = "2017"
}

@article{Iuliano:2024ogr,
    author = "Iuliano, Claudio and Hollands, Stefan and Green, Stephen R. and Zimmerman, Peter",
    title = "{Extremal black hole weather}",
    eprint = "2412.02821",
    archivePrefix = "arXiv",
    primaryClass = "gr-qc",
    doi = "10.1103/PhysRevD.111.124038",
    journal = "Phys. Rev. D",
    volume = "111",
    number = "12",
    pages = "124038",
    year = "2025"
}

@article{Siemonsen:2025fne,
    author = "Siemonsen, Nils",
    title = "{Weakly turbulent saturation of the nonlinear scalar ergoregion instability}",
    eprint = "2510.07467",
    archivePrefix = "arXiv",
    primaryClass = "gr-qc",
    month = "10",
    year = "2025"
}

@article{Krynicki:2025fzi,
    author = "Krynicki, Holly and Wu, Jiaxi and Most, Elias R.",
    title = "{Toward a Theory of Gravitational Wave Turbulence}",
    eprint = "2509.19769",
    archivePrefix = "arXiv",
    primaryClass = "gr-qc",
    month = "9",
    year = "2025"
}

@article{Benomio:2024lev,
    author = "Benomio, Gabriele and C{\'a}rdenas-Avenda{\~n}o, Alejandro and Pretorius, Frans and Sullivan, Andrew",
    title = "{Turbulence for spacetimes with stable trapping}",
    eprint = "2411.17445",
    archivePrefix = "arXiv",
    primaryClass = "gr-qc",
    doi = "10.1103/PhysRevD.111.104037",
    journal = "Phys. Rev. D",
    volume = "111",
    number = "10",
    pages = "104037",
    year = "2025"
}

@article{Galtier:2018vbq,
    author = "Galtier, S{\'e}bastien and Nazarenko, Sergey V. and Buchlin, {\'E}ric and Thalabard, Simon",
    title = "{Nonlinear Diffusion Models for Gravitational Wave Turbulence}",
    eprint = "1809.07623",
    archivePrefix = "arXiv",
    primaryClass = "gr-qc",
    doi = "10.1016/j.physd.2019.01.007",
    journal = "Physica D",
    volume = "390",
    pages = "84--88",
    year = "2019"
}

@article{Yang:2014tla,
    author = "Yang, Huan and Zimmerman, Aaron and Lehner, Luis",
    title = "{Turbulent Black Holes}",
    eprint = "1402.4859",
    archivePrefix = "arXiv",
    primaryClass = "gr-qc",
    doi = "10.1103/PhysRevLett.114.081101",
    journal = "Phys. Rev. Lett.",
    volume = "114",
    pages = "081101",
    year = "2015"
}

@article{Fransen:2025cgv,
    author = "Fransen, Kwinten and Pere{\~n}iguez, David and Redondo-Yuste, Jaime",
    title = "{Perturbations of Plane Waves and Quadratic Quasinormal Modes on the Lightring}",
    eprint = "2509.03598",
    archivePrefix = "arXiv",
    primaryClass = "gr-qc",
    month = "9",
    year = "2025"
}

@article{BenAchour:2024skv,
    author = "Ben Achour, Jibril and Roussille, Hugo",
    title = "{Quadratic perturbations of the Schwarzschild black hole: the algebraically special sector}",
    eprint = "2406.08159",
    archivePrefix = "arXiv",
    primaryClass = "gr-qc",
    doi = "10.1088/1475-7516/2024/07/085",
    journal = "JCAP",
    volume = "07",
    pages = "085",
    year = "2024"
}

@article{Ma:2025rnv,
    author = "Ma, Sizheng and Lehner, Luis and Yang, Huan and Kidder, Lawrence E. and Pfeiffer, Harald P. and Scheel, Mark A.",
    title = "{Emergent Turbulence in Nonlinear Gravity}",
    eprint = "2508.13294",
    archivePrefix = "arXiv",
    primaryClass = "gr-qc",
    month = "8",
    year = "2025"
}

@article{Ma:2021znq,
    author = "Ma, Sizheng and Giesler, Matthew and Varma, Vijay and Scheel, Mark A. and Chen, Yanbei",
    title = "{Universal features of gravitational waves emitted by superkick binary black hole systems}",
    eprint = "2107.04890",
    archivePrefix = "arXiv",
    primaryClass = "gr-qc",
    doi = "10.1103/PhysRevD.104.084003",
    journal = "Phys. Rev. D",
    volume = "104",
    number = "8",
    pages = "084003",
    year = "2021"
}

@article{Ma:2022wpv,
    author = "Ma, Sizheng and Mitman, Keefe and Sun, Ling and Deppe, Nils and H\'ebert, Fran\c{c}ois and Kidder, Lawrence E. and Moxon, Jordan and Throwe, William and Vu, Nils L. and Chen, Yanbei",
    title = "{Quasinormal-mode filters: A new approach to analyze the gravitational-wave ringdown of binary black-hole mergers}",
    eprint = "2207.10870",
    archivePrefix = "arXiv",
    primaryClass = "gr-qc",
    doi = "10.1103/PhysRevD.106.084036",
    journal = "Phys. Rev. D",
    volume = "106",
    number = "8",
    pages = "084036",
    year = "2022"
}

@article{Bourg:2024jme,
    author = "Bourg, Patrick and Panosso Macedo, Rodrigo and Spiers, Andrew and Leather, Benjamin and Bonga, B\'eatrice and Pound, Adam",
    title = "{Quadratic quasi-normal mode dependence on linear mode parity}",
    eprint = "2405.10270",
    archivePrefix = "arXiv",
    primaryClass = "gr-qc",
    month = "5",
    year = "2024"
}

@article{Ma:2024qcv,
    author = "Ma, Sizheng and Yang, Huan",
    title = "{Excitation of quadratic quasinormal modes for Kerr black holes}",
    eprint = "2401.15516",
    archivePrefix = "arXiv",
    primaryClass = "gr-qc",
    doi = "10.1103/PhysRevD.109.104070",
    journal = "Phys. Rev. D",
    volume = "109",
    number = "10",
    pages = "104070",
    year = "2024"
}

@article{Cheung:2023vki,
    author = "Cheung, Mark Ho-Yeuk and Berti, Emanuele and Baibhav, Vishal and Cotesta, Roberto",
    title = "{Extracting linear and nonlinear quasinormal modes from black hole merger simulations}",
    eprint = "2310.04489",
    archivePrefix = "arXiv",
    primaryClass = "gr-qc",
    doi = "10.1103/PhysRevD.109.044069",
    journal = "Phys. Rev. D",
    volume = "109",
    number = "4",
    pages = "044069",
    year = "2024",
    note = "[Erratum: Phys.Rev.D 110, 049902 (2024)]"
}

@article{Siegel:2023lxl,
    author = "Siegel, Harrison and Isi, Maximiliano and Farr, Will M.",
    title = "{Ringdown of GW190521: Hints of multiple quasinormal modes with a precessional interpretation}",
    eprint = "2307.11975",
    archivePrefix = "arXiv",
    primaryClass = "gr-qc",
    reportNumber = "LIGO-P2300214",
    doi = "10.1103/PhysRevD.108.064008",
    journal = "Phys. Rev. D",
    volume = "108",
    number = "6",
    pages = "064008",
    year = "2023"
}

@article{Mitman:2022qdl,
    author = "Mitman, Keefe and others",
    title = "{Nonlinearities in Black Hole Ringdowns}",
    eprint = "2208.07380",
    archivePrefix = "arXiv",
    primaryClass = "gr-qc",
    doi = "10.1103/PhysRevLett.130.081402",
    journal = "Phys. Rev. Lett.",
    volume = "130",
    number = "8",
    pages = "081402",
    year = "2023"
}

@article{Isi:2022mhy,
    author = "Isi, Maximiliano and Farr, Will M.",
    title = "{Revisiting the ringdown of GW150914}",
    eprint = "2202.02941",
    archivePrefix = "arXiv",
    primaryClass = "gr-qc",
    reportNumber = "LIGO-P2200028",
    month = "2",
    year = "2022"
}

@article{Cotesta:2022pci,
    author = "Cotesta, Roberto and Carullo, Gregorio and Berti, Emanuele and Cardoso, Vitor",
    title = "{Analysis of Ringdown Overtones in GW150914}",
    eprint = "2201.00822",
    archivePrefix = "arXiv",
    primaryClass = "gr-qc",
    doi = "10.1103/PhysRevLett.129.111102",
    journal = "Phys. Rev. Lett.",
    volume = "129",
    number = "11",
    pages = "111102",
    year = "2022"
}

@article{Ota:2019bzl,
    author = "Ota, Iara and Chirenti, Cecilia",
    title = "{Overtones or higher harmonics? Prospects for testing the no-hair theorem with gravitational wave detections}",
    eprint = "1911.00440",
    archivePrefix = "arXiv",
    primaryClass = "gr-qc",
    doi = "10.1103/PhysRevD.101.104005",
    journal = "Phys. Rev. D",
    volume = "101",
    number = "10",
    pages = "104005",
    year = "2020"
}

@article{Bhagwat:2019dtm,
    author = "Bhagwat, Swetha and Forteza, Xisco Jimenez and Pani, Paolo and Ferrari, Valeria",
    title = "{Ringdown overtones, black hole spectroscopy, and no-hair theorem tests}",
    eprint = "1910.08708",
    archivePrefix = "arXiv",
    primaryClass = "gr-qc",
    doi = "10.1103/PhysRevD.101.044033",
    journal = "Phys. Rev. D",
    volume = "101",
    number = "4",
    pages = "044033",
    year = "2020"
}

@article{Berti:2005ys,
    author = "Berti, Emanuele and Cardoso, Vitor and Will, Clifford M.",
    title = "{On gravitational-wave spectroscopy of massive black holes with the space interferometer LISA}",
    eprint = "gr-qc/0512160",
    archivePrefix = "arXiv",
    doi = "10.1103/PhysRevD.73.064030",
    journal = "Phys. Rev. D",
    volume = "73",
    pages = "064030",
    year = "2006"
}

@article{Kokkotas:1999bd,
    author = "Kokkotas, Kostas D. and Schmidt, Bernd G.",
    title = "{Quasinormal modes of stars and black holes}",
    eprint = "gr-qc/9909058",
    archivePrefix = "arXiv",
    doi = "10.12942/lrr-1999-2",
    journal = "Living Rev. Rel.",
    volume = "2",
    pages = "2",
    year = "1999"
}

@misc{bourg2025quadraticquasinormalmodesnull,
      title={Quadratic quasinormal modes at null infinity on a Schwarzschild spacetime}, 
      author={Patrick Bourg and Rodrigo Panosso Macedo and Andrew Spiers and Benjamin Leather and Bonga Béatrice and Adam Pound},
      year={2025},
      eprint={2503.07432},
      archivePrefix={arXiv},
      primaryClass={gr-qc},
      url={https://arxiv.org/abs/2503.07432}, 
}

@book{Boyd:2008eba,
    author = "Boyd, Robert W.",
    title = "{Nonlinear Optics}",
    isbn = "978-0-12-369470-6",
    publisher = "Academic Press",
    address = "Burlington, MA",
    year = "2008"
}

@book{SalehTeich2007,
  author    = {Saleh, Bahaa E. A. and Teich, Malvin C.},
  title     = {Fundamentals of Photonics},
  edition   = {2},
  year      = {2007},
  publisher = {Wiley},
  address   = {Hoboken, NJ},
  isbn      = {978-0-471-35832-9}
}

@article{Green:2022htq,
    author = "Green, Stephen R. and Hollands, Stefan and Sberna, Laura and Toomani, Vahid and Zimmerman, Peter",
    title = "{Conserved currents for a Kerr black hole and orthogonality of quasinormal modes}",
    eprint = "2210.15935",
    archivePrefix = "arXiv",
    primaryClass = "gr-qc",
   doi ="10.1103/PhysRevD.107.064030",
   journal = "Phys. Rev. D",
    volume = "107",
    number = "6",
    pages = "064030",
    year = "2023"
}

@article{Cannizzaro:2023jle,
    author = "Cannizzaro, Enrico and Sberna, Laura and Green, Stephen R. and Hollands, Stefan",
    title = "{Relativistic Perturbation Theory for Black-Hole Boson Clouds}",
    eprint = "2309.10021",
    archivePrefix = "arXiv",
    primaryClass = "gr-qc",
    doi = "10.1103/PhysRevLett.132.051401",
    journal = "Phys. Rev. Lett.",
    volume = "132",
    number = "5",
    pages = "051401",
    year = "2024"
}

@book{Nayfeh1973,
  author    = {Nayfeh, Ali H.},
  title     = {Perturbation Methods},
  publisher = {Wiley-Interscience},
  address   = {New York},
  year      = {1973},
  isbn      = {978-0471359333}
}

\end{document}